\documentstyle[multicol,aps]{revtex}

\begin{document}

\draft

\title
{
Number of Transition Frequencies of a System Containing an Arbitrary 
Number of Gas Bubbles
}

\author
{ 
Masato Ida
}

\address
{
Satellite Venture Business Laboratory, Gunma University, 1--5--1 Tenjin-cho, Kiryu-shi, Gunma 376-8515, Japan
}

\maketitle

\begin{abstract}
``Transition frequencies'' of a system containing an arbitrary number of 
bubbles levitated in a liquid are discussed. Using a linear coupled-oscillator 
model, it is shown theoretically that when the system contains $N$ bubbles of 
different sizes, each bubble has $2N - 1$ (or less) transition frequencies 
which make the phase difference between an external sound and a bubble's 
pulsation $\pi / 2$. Furthermore, we discuss a discrepancy appearing between 
the present result regarding the transition frequencies and existing 
ones for the resonance frequencies in a two-bubble case, and show that the 
transition frequency, defined as above, and the resonance frequency have a 
different physical meaning when $N \ge 2$, while they are consistent for 
$N = 1$.
\end{abstract}

\begin{multicols}{2}

Multibubble dynamics in an acoustic field have been studied in a variety of 
fields \cite{ref2,ref3,ref4,ref5,ref6,ref7,ref9,ref11,ref12,ref13,ref15,ref16,ref1}, and the mutual radiative interaction of bubbles is 
known to change their acoustic properties 
\cite{ref2,ref3,ref4,ref5,ref6,ref7,ref11,ref12,ref16}. 
Shima \cite{ref2} derived a theoretical formula for the resonance 
frequencies of two mutually interacting gas bubbles levitated in an inviscid 
liquid, and showed that a bubble interacting with a neighboring bubble has two 
resonance frequencies, which reveal an upward or downward shift as the bubbles 
approach each other. Zabolotskaya \cite{ref3} showed theoretically that 
variations in the resonance 
frequencies may cause changes of phase difference between bubbles, and 
may, in turn, sometimes give rise to reversal of the sign of the 
secondary Bjerknes force, which is an interaction force acting between 
pulsating bubbles \cite{ref13}. 
In the studies described in refs.~3 and 4, Doinikov and Zavtrak 
produced results that were almost equivalent to those of Zabolotskaya by 
employing a mathematical model in which the multiple scattering between bubbles 
is described using a Legendre expansion function. Their results are considered 
to explain the ``bubble grape'', which is a stable structure formed by 
bubbles in an acoustic field \cite{ref5}. Feuillade \cite{ref6}, using a 
self-consistent model, showed that when two bubbles are pulsating in antiphase 
and radiation damping is dominant rather than thermal or viscous damping, 
the bubbles may show a superresonant response \cite{ref7} because of the 
reduction in the effective damping. Ye and Alvarez \cite{ref9}, using a 
self-consistent model, 
carried out numerical experiments on acoustic wave propagation in a liquid 
containing randomly distributed identical bubbles, and showed that the 
localization of acoustic waves can take place in a range of frequencies. It 
was considered that the multiple scattering of sound due to the bubbles has 
a strong influence on this phenomenon. Mettin \textit{et al}.~\cite{ref11} and 
Doinikov \cite{ref12} examined some influences of the interaction of two 
bubbles in a strong acoustic field on the radiation forces (the secondary 
\cite{ref13} and primary \cite{ref17,ref14} Bjerknes forces) acting on the 
bubbles; these studies are related to multibubble sonoluminescence 
\cite{add3,add1}.

The aim of this paper is to provide preliminary but significant discussions 
regarding the ``transition frequency'' of a system containing an arbitrary 
number of bubbles having different radii, by employing a simple theoretical 
model. In the present study, the transition frequencies of a bubble in the 
system are defined as \textit{the frequencies of an external sound for which 
the phase difference between the sound and the bubble's pulsation becomes} 
${\it \pi / 2}$ \cite{ref16}. This definition is based on the well-known 
phenomenon of a single bubble case in which the phase difference between an 
external sound and the bubble pulsation becomes $\pi / 2$ at its resonance 
frequency (more correctly, at its natural frequency) \cite{ref1,ref18}. This 
study has two main purposes: (1) To determine how many transition frequencies a 
bubble in the system has. As has been reported, it can be considered that the 
phase shifts of the bubble pulsations due to the radiation coupling give rise 
to a large variety of phenomena that cannot be predicted by a single-bubble 
model (see, e.g., refs.~1--9). However, 
there exist only a few studies on the transition frequencies defined as 
mentioned above. Almost all of the previous works (e.g., refs.~1--5) concern 
the resonance frequencies. Very 
recently, Ida \cite{ref16}, using a coupled oscillator model, showed 
theoretically that a bubble interacting with a neighboring bubble has three 
transition frequencies. In the present paper, by extending this approach, we 
derive the number of transition frequencies of a bubble interacting with an 
arbitrary number of bubbles. (2) To elucidate a difference between the 
transition frequency and the resonance frequency. As suggested already, the 
transition frequency and the resonance frequency have different physical 
meanings when the number of bubbles, $N$, is larger than 1. As briefly reviewed 
above, Shima\cite{ref2} and Zabolotskaya\cite{ref3} predicted the existence of 
two resonance frequencies per bubble when $N = 2$. This number of resonance 
frequencies is not equal to that of the transition frequencies given by 
Ida\cite{ref16}. In the 
present paper, we show that in general, the numbers of those frequencies are 
different. This result shows that when $N \ge 2$, the reversal of the phase of 
the bubble pulsation (e.g., from in-phase to out-of-phase with an external 
sound) may occur not only at the resonance but also at other frequencies, 
unlike the case of $N = 1$.

At first, we derive a system of linear differential equations that describes 
the pulsation of coupled bubbles and that is almost equivalent to those used in 
refs.~1, 2, 5, 12 and 19. Let us assume that $N$ bubbles are levitated in a 
viscous liquid and the equilibrium radius of each bubble (assumed to be 
constant \cite{add12}) is $R_{i0}$ ($i = 1,2, \cdots ,N$). When the 
amplitude of the external sound is weak, the bubble 
pulsation can be described by a linear second-order differential equation 
\cite{add2,ref19,ref1}
\begin{equation}
\label{eq1}
\ddot e_i + \omega _{i0}^2 e_i + \delta _i \dot e_i 
= - \frac{1}{\rho R_{i0}}p_{{\rm d}\,i} ,
\end{equation}
where it is assumed that the sphericity of the bubbles is maintained and the 
time-dependent bubble radii $R_i (t)$ can be represented by 
$R_i (t) = R_{i0} + e_i (t)$ ($\left| {e_i } \right| \ll R_{i0}$), and 
$\omega _{i0}$ ($ = \sqrt {[3\kappa P_0  + (3\kappa  - 1)2\sigma /R_{i0} ] 
/ \rho R_{i0}^2 }$) is the resonance (angular) frequency, $\delta _i$ is the 
damping factor \cite{add11}, $p_{{\rm d}\,i}$ is the driving pressure acting 
on bubble $i$, $\rho$ is the density of the surrounding liquid, $\kappa$ is the 
polytropic exponent of the gas inside the bubbles, $P_0 $ is the equilibrium 
pressure, $\sigma$ is the surface-tension coefficient, and the over dots denote 
the derivation with respect to time. When more than one bubble exists, the 
driving pressure $p_{{\rm d}\,i}$ is given by 
the sum of the pressures of the external sound, $p_{ex}$, and the scattered 
sound waves due to other bubbles, $p_{{\rm s}\,ij}$, as \cite{add10}
\begin{equation}
\label{eq2}
p_{{\rm d}\,i} = p_{{\rm ex}} + \sum\limits_{j = 1,\;j \ne i}^N 
{p_{{\rm s}\,ij} } \approx p_{\rm ex} + \sum\limits_{j = 1,\;j \ne i}^N 
{\frac{\rho R_{j0}^2}{D_{ij} }\ddot e_j} ,
\end{equation}
where the surrounding liquid is assumed to be incompressible, and $D_{ij}$ is 
the distance between the centers of bubbles $i$ and $j$. By substituting 
eq.~(\ref{eq2}) into eq.~(\ref{eq1}), one obtains
\begin{equation}
\label{eq3}
\ddot e_i  + \omega _{i0}^2 e_i + \delta _i \dot e_i = - \frac{{p_{{\rm ex}} }}{{\rho R_{i0} }} - \frac{1}{{R_{i0} }}\sum\limits_{j = 1,\;j \ne i}^N {\frac{{R_{j0}^2 }}{{D_{ij} }}\ddot e_j } .
\end{equation}
This kind of system of differential equations is called a coupled-oscillator 
model or a self-consistent model. Now we assume that $p_{{\rm ex}} = 
P\exp ({\rm i}\,\omega t)$ and $e_i = A_i \exp({\rm i}\,\omega t)$, 
where $P$ is a real, positive constant and $A_i$ denotes complex amplitudes. 
Those assumptions reduce eq.~(\ref{eq3}) to
\begin{equation}
\label{eq4}
R_{i0} [(\omega ^2  - \omega _{i0}^2 ) - {\rm i}\,\omega \delta _i ]\,A_i 
+ \sum\limits_{j = 1,\;j \ne i}^N {\frac{R_{j0}^2}{D_{ij}}\omega ^2 A_j } 
= \frac{P}{\rho } .
\end{equation}
This system of equations can be written in a matrix form,
\begin{equation}
\label{eq5}
{\bf MA} = {\bf B} ,
\end{equation}
where ${\bf M}$ is an $N \times N$ matrix whose elements, $m_{ij}$, are 
defined as
\begin{equation}
\label{eq6}
m_{ij}  = \left\{ {\begin{array}{*{20}c}
{R_{i0} [(\omega ^2  - \omega _{i0}^2 ) - {\rm i}\omega \delta _i ]}, \hfill 
\quad & {{\rm for}\;\;i = j,} \\
{\displaystyle \frac{R_{j0}^2}{D_{ij}}\omega ^2}, \hfill \quad & 
{{\rm otherwise,}} \\
\end{array}} \right.
\end{equation}
${\bf A} = (A_1 ,A_2 , \cdots ,A_N )^T$ and ${\bf B} = (P / \rho) 
(1,1, \cdots ,1)^T$.

Let us analyze the matrix equation. The solution of eq.~(\ref{eq5}) is 
expressed as
\begin{equation}
\label{eq7}
{\bf A} = {\bf M}^{ - 1} {\bf B} .
\end{equation}
As is well known, ${\bf M}^{ - 1}$ can be written in the form of
\begin{equation}
\label{eq8}
{\bf M}^{ - 1}  = \frac{{\bf C}}{{\left| {\bf M} \right|}},
\end{equation}
where $\left| {\bf M} \right|$ and ${\bf C}$ are the determinant 
and the cofactor matrix, respectively, of ${\bf M}$. From the definitions of 
the determinant and the cofactor of a matrix, one determines that
\begin{eqnarray}
\label{eq9}
&\deg \left( {\left| {{\bf M}(\omega )} \right|} \right) = 2N ,& \\
\label{eq10}
&\deg \left( {{\bf C}(\omega )} \right) = 2(N - 1) .&
\end{eqnarray}
Here we should note that $\deg \left( {m_{ij} (\omega )} \right) = 2$. 
Substitution of eq.~(\ref{eq8}) into eq.~(\ref{eq7}) yields
\[
{\bf A} = \frac{1}{{\left| {\bf M} \right|}}{\bf C}\,{\bf B} .
\]
As defined previously, the transition frequencies of bubble $i$ are derived by 
an equation of
\begin{equation}
\label{eq11}
{\mathop{\rm Re}\nolimits} (A_i ) = 0 .
\end{equation}
Assuming that
\begin{eqnarray}
\label{eq12}
&\left| {\bf M} \right| = a + {\rm i}\,b ,& \\
\label{eq13}
&g_i = c_i + {\rm i}\,d_i ,&
\end{eqnarray}
where $a$, $b$, $c_i$, and $d_i$ are real values and $g_i$ is an element of 
${\bf C}\,{\bf B}$, yields
\[
A_i = \frac{c_i + {\rm i}\,d_i}{a + {\rm i}\,b} 
= \frac{ac_i + bd_i + {\rm i}\,(ad_i  - bc_i )}{a^2  + b^2}
\]
and
\begin{equation}
\label{eq14}
{\mathop{\rm Re}\nolimits} (A_i ) = \frac{{ac_i  + bd_i }}{{a^2  + b^2 }} .
\end{equation}
Using this equation and assuming that $\left| {\bf M} \right| \ne 0$, i.e., 
no case exists where both $a = 0$ and $b = 0$ are true (this assumption is 
physically valid because this guarantees the absence of a singular solution), 
eq.~(\ref{eq11}) is reduced to
\begin{equation}
\label{eq15}
ac_i  + bd_i = 0 .
\end{equation}
Now we examine how many roots this equation has. From eqs.~(\ref{eq9}) and 
(\ref{eq10}) and definitions (\ref{eq12}) and (\ref{eq13}), we know that 
$\deg (ac_i  + bd_i ) = 2N + 2(N - 1) = 4N - 2$. Furthermore, using definition 
(\ref{eq6}), we can prove that eq.~(\ref{eq15}) contains terms of only even 
orders with respect to $\omega$ as follows: the real parts of $m_{ij}$ are of 
the second order and the imaginary ones are of the 
first order with respect to $\omega$; thus, only terms of even orders remain in 
the real part of $A_i$, because an even-order component given by the imaginary 
parts of $m_{ij}$, e.g., ${\rm i}\,\omega \delta _1 \times {\rm i}\,\omega 
\delta _2 = - \delta _1 \delta _2 \omega ^2$, becomes a real value and an 
odd-order component, e.g., ${\rm i}\,\omega \delta _1 \times {\rm i}\,\omega 
\delta _2 \times {\rm i}\,\omega \delta _3 = - {\rm i}\,\delta _1 \delta _2 
\delta _3 \omega ^3$, becomes an imaginary value. As a result, eq.~(\ref{eq15}) 
can be represented by
\begin{equation}
\label{eq16}
F(X) \equiv ac_i + bd_i = 0 ,
\end{equation}
and one knows that
\begin{equation}
\label{eq17}
\deg \left( {F(X)} \right) = 2N - 1 ,
\end{equation}
with $X = \omega ^2$. These equations predict that a bubble has $2N - 1$ 
(or less) transition 
frequencies. (Note that only a positive root is physical.) When $N = 2$, for 
example, $2N - 1 = 3$; this result is in agreement with that given by Ida 
\cite{ref16}.

To confirm the above result, we show here a numerical result for $N = 3$. 
The parameters used are $R_{10} = 1$ $\mu$m, $R_{20} = 1.5$ $\mu$m, 
$R_{30} = 2.5$ $\mu$m, 
$\rho = 1000$ kg/m$^3$, $\kappa = 1.4$, $P_0 = 1$ atm, and $\sigma = 0.0728$ 
N/m. The damping factor is set to $\delta _i  = 4\mu /\rho R_{i0}^2$, 
where $\mu$ ($= 1.137 \times 10^{- 3}$ kg/(m s)) is the viscosity of water 
at room temperature (i.e., the viscous damping is adapted, which is dominant 
for small bubbles used, e.g., in medical applications \cite{ref15} and 
in the experiments on sonoluminescence \cite{add3,ref1}), and $D_{ij}$ is 
determined by $D_{ij} = s (R_{i0} + R_{j0})$. 
For $s = 2.0$, we, as expected, obtain five ($= 2 \times 3 - 1$) transition 
frequencies of bubble 1, $\omega _1 = (1.054, 0.723, 0.635, 0.411, 0.328) 
\times \omega _{10}$, while for $s = 10.0$, only one, $\omega _1 = 
1.002\omega _{10}$, is found. Such a dependency of the number on the distance 
between bubbles can also be found in a two-bubble case \cite{ref16}. Detailed 
discussion regarding three- or more-bubble cases will be given in a future 
paper.

We perform here a comparative study between the present theory and previous 
ones in order to clarify a difference between the transition frequency and the 
resonance frequency. Shima \cite{ref2} and Zabolotskaya \cite{ref3} derived the 
same formula for estimating the resonance frequencies in the case of $N = 2$, 
expressed as
\begin{equation}
\label{eq18}
(X - \omega _{10}^2 )(X - \omega _{20}^2 ) - \frac{{R_{10} R_{20}}}{{D_{21}^2 }}X^2 = 0 .
\end{equation}
In their mathematical models, the sphericity of bubbles and 
incompressibility of the surrounding liquid were assumed as in the 
present one, but the damping factor was neglected. This formula predicts two 
resonance frequencies; apparently, this result seems to contradict the 
result given by Ida, which predicts three transition frequencies\cite{ref16}. 
We next discuss the origin of this discrepancy.

Generally, when no damping factor exists ($\delta _i = 0$, 
${\mathop{\rm Im}\nolimits} (m_{ij} ) = 0$), the resonance frequencies are 
determined by
\begin{equation}
\label{eq19}
\left| {\bf M} \right| = 0
\end{equation}
in the case of this study. This condition gives rise to the well-known 
singularity in the bubble pulsation at the resonance frequencies (infinite 
amplitudes and discontinuous phase reversals) \cite{ref20,ref17}. This 
formula is reduced to the well-known one for a single-bubble case, 
$\omega ^2 - \omega _{10}^2  = 0$, and to eq.~(\ref{eq18}) for a double-bubble 
case. On the other hand, in the present study, the transition frequencies of 
bubble $i$ in the case of $\delta _i = 0$ are derived by the following 
independent equations:
\begin{equation}
\label{eq20}
{\mathop{\rm Re}\nolimits} (A_i ) = A_i = \frac{{g_i }}{{\left| {\bf M} \right|}} = 0
\end{equation}
and
\begin{equation}
\label{eq21}
{\rm Sign}\left( {\mathop {\lim }\limits_{\Delta  \to  - 0} A_i (\omega  + \Delta )} \right) \ne {\rm Sign}\left( {\mathop {\lim }\limits_{\Delta  \to  + 0} A_i (\omega  + \Delta )} \right) ,
\end{equation}
where
\[
{\rm Sign}\left( f \right) = \left\{ {\begin{array}{*{20}c}
1 \hfill & {{\rm for}\;\;f > 0,} \hfill  \\
{ - 1} \hfill & {{\rm for}\;\;f < 0.} \hfill  \\
\end{array}} \right.
\]
Equation (\ref{eq21}) means that the phase of bubble $i$ is reversed at 
$\omega $ satisfying this equation. 
(In the limit of $\delta _i \to 0$, the condition that the phase difference 
is $\pi /2$ converges to eq.~(\ref{eq21}) at the resonances because of the 
singularity mentioned above. Namely, at the resonances those conditions are 
essentially equivalent.) Equations (\ref{eq20}) and (\ref{eq21}), respectively, 
may be reduced to $g_i = 0$ and $\left| {\bf M} \right| = 0$, or
\begin{equation}
\label{eq22}
\left| {\bf M} \right| g_i = 0 .
\end{equation}
Furthermore, using the definitions of eqs.~(\ref{eq12}) and (\ref{eq13}), 
eq.~(\ref{eq22}) can be reduced to
\[
ac_i = 0 .
\]
This equation is in agreement with eq.~(\ref{eq15}) when $b = 0$ and $d_i = 0$ 
(these conditions are naturally satisfied because of 
${\mathop{\rm Im}\nolimits} (m_{ij} ) = 0$). The degree of eq.~(\ref{eq19}) 
with respect to $X$ is $N$ as 
discussed above, while that of eq.~(\ref{eq22}) is $N + (N - 1) = 2N - 1$. 
When $N > 1$, these two degrees are not equal to one another. This result 
reveals that (1) ``resonance frequency'' and ``transition frequency'' have 
different physical meanings, (2) the number of the transition frequencies is, 
in general, larger than that of the resonance frequency, and (3) other 
conditions under which the phase reversal of bubble pulsation takes place exist 
in addition to those for the resonance. The transition frequencies given 
by $\left| {\bf M} \right| = 0$ cause the resonance response of the bubbles, 
while those given by $g_i = 0$ do not cause it. (The respective transition 
frequencies may correspond to the series and the parallel resonance frequencies 
of an electromagnetic circuit, which give rise to a zero and an infinite 
impedance, respectively.)

Next, we discuss the case of $D_{ij} \approx \infty $ and 
$\delta _i \approx 0$, i.e., in which the non-diagonal elements of ${\bf M}$ 
and the imaginary parts of its diagonal elements have quite small absolute 
values. These settings result in
\[
\begin{array}{*{20}c}
{a = \prod\limits_{j = 1}^N {R_{j0} (\omega ^2 - \omega _{j0}^2 )} 
+ \varepsilon 1,} \hfill \quad & {b = \varepsilon 2,} \hfill  \\
{{\displaystyle c_i = \frac{P}{\rho }\prod\limits_{j = 1,\,j \ne i}^N 
{R_{j0} (\omega ^2 - \omega _{j0}^2 )} + \varepsilon 3,}} \quad & {d_i = \varepsilon 4,} \hfill \\
\end{array}
\]
where $\varepsilon 1 \sim \varepsilon 4$ are real 
values whose absolute values are quite small. They reduce eq.~(\ref{eq15}) to
\[
\left( {\prod\limits_{j = 1}^N {R_{j0} (\omega ^2 - \omega _{j0}^2 )} } \right)
\left( {\frac{P}{\rho }\prod\limits_{j = 1,\,j \ne i}^N {R_{j0} 
(\omega ^2 - \omega _{j0}^2 )} } \right) \approx 0
\]
or
\[
(\omega ^2  - \omega _{i0}^2 )\prod\limits_{j = 1,\,j \ne i}^N 
{(\omega ^2 - \omega _{j0}^2 )^2 } \approx 0 .
\]
This equation shows that for $D_{ij} \to \infty$, among $2N - 1$ transition 
frequencies of bubble $i$, only one converges to $\omega _{i0}$, while two of 
the remaining ones converge to $\omega _{j0}$ ($j \ne i$). This result is 
consistent with that for $N = 2$. \cite{ref16}

Lastly, we briefly discuss a special case. Here we assume that all bubbles 
are identical ($R_{10} = R_{20} = \cdots = R_{N0}$, 
$\delta _1 = \delta _2 = \cdots = \delta _N $) and, furthermore, that 
the distances between all pairs of bubbles are uniform. (The latter 
assumption may be realizable up to $N = 4$.) When $N = 3$, this setting 
corresponds to the ``mode $A$'' discussed in Sec.~III of ref.~5. 
Based on these assumptions, the elements of matrix $\left| {\bf M} \right|$ 
become
\begin{eqnarray*}
&m_{11} = m_{22} = \cdots = m_{NN} = R_{10} [(\omega ^2 - \omega _{10}^2 ) 
- {\rm i}\,\omega \delta _1 ] , & \\
&{\displaystyle m_{12} = m_{21} = m_{13} = \cdots m_{NN - 1} = 
\frac{{R_{10}^2 }}{{D_{12} }}\omega ^2} . &
\end{eqnarray*}
Namely, both the diagonal and the non-diagonal elements become uniform. As a 
result, one obtains $A_1 = A_2 = \cdots = A_N$ and 
\[
A_1 = \frac{1}{{m_{11} + (N - 1)m_{12} }} \cdot \frac{P}{\rho } .
\]
Only one transition frequency (corresponding to the resonance frequency) is 
given by this formula, although $2N - 1$ ones are predicted in general. This 
result points out that the numbers of both the resonance and the transition 
frequencies are reduced under certain conditions and 
that such an excessive idealization prevents the accurate understanding of 
multibubble dynamics in an acoustic field.

In summary, we have studied the transition frequencies of a system containing 
an arbitrary number ($N$) of bubbles of different sizes by employing a linear 
coupled-oscillator model. In the present study, the frequency of an external 
sound in the case where the phase difference between the external sound and a 
bubble's pulsation becomes $\pi /2$ was called the transition frequency. The 
present theory predicts that a bubble in the system has $2N - 1$ transition 
frequencies, while it has $N$ resonance frequencies; namely, the well-known 
theory for a one-bubble case in which the phase reversal of a bubble's 
pulsation takes place around its resonance frequency is not absolutely true in 
a two- or more-bubble case. 
(The physical meaning of the existence of an odd number of transition 
frequencies may be explained as follows: a bubble, even if interacting with 
neighboring bubbles, pulsates in-phase or out-of-phase with an external sound 
if the driving frequency is much lower or much higher, respectively, than the 
resonance frequencies of the bubbles; to interpolate those two extreme states 
consistently, an odd number of the phase reversals is needed.) 
This result is important especially for subjects in 
which the phase change of bubbles plays an important role, such as those 
studied in refs.~2--4, 8 and 9 and investigated using only the resonance 
frequencies. (In ref.~26, the author already applied the theory for the 
transition frequency to the investigation of the reversal of the sign of the 
secondary Bjerknes force, and found that the theory gives an 
alternative interpretation of the phenomenon, which may be more accurate than 
the previous ones given in refs.~2--4.) 
We expect that more detailed mathematical discussions regarding the matrix 
equation will supply a still richer understanding of the physics of multibubble 
dynamics in an acoustic field and acoustic wave propagation in bubbly liquid.

\end{multicols}


\begin{thebibliography}{99}
\bibitem{add11}Generally, $\delta _i$ is determined by the sum of the viscous, 
acoustic and thermal damping coefficients, the second and the last ones of 
which depend on the driving frequency \cite{add2,ref6}. The present paper, 
however, treats $\delta _i$ as being independent of the frequency, to simplify 
the discussion. The numerical results regarding two-bubble cases \cite{add4} 
reveal that even if the acoustic and thermal damping are taken into account, 
the maximum number of the transition frequencies does not change.
\bibitem{add10}The amplitude of the scattered wave can be given, e.g., by 
integrating the momentum equation for linear sound waves, 
$\partial p/\partial r =  - \rho \partial u/\partial t$, coupled with the 
divergence-free condition, $\partial (r^2 u)/\partial r = 0$, where $r$ is the 
radial coordinate measured from the center of a bubble and $u$ is the velocity 
along $r$.
\end{thebibliography}
\end{document}